\documentclass[fleqn,11pt]{article}
\usepackage{amsmath,amssymb}
\usepackage{xcolor}
\usepackage{verbatim}
\usepackage{amsfonts,amssymb,cite}
\usepackage{graphicx}
\def\noi{\noindent}

\newcommand{\Title}[1]{\noi {{\Large\bf #1}}\\[1ex]}

\newcommand{\Author}[2]{\noi{\bf #1}\\[2ex]\noi{\normalsize\it #2}\\}

\newcommand{\Abstract}[1]{\vskip 2mm \begin{center}
        \parbox{16.4cm}{\small\noi #1} \end{center}\medskip}
\newcommand{\foom}[1]{\protect\footnotemark[#1]}

\def\nqq{\hspace*{-2em}}

\def\Jl#1#2{#1 {\bf #2},\ }

\def\ApJ#1 {\Jl{Astroph. J.}{#1}}
\def\CQG#1 {\Jl{Class. Quantum Grav.}{#1}}
\def\DAN#1 {\Jl{Dokl. AN SSSR}{#1}}
\def\GC#1 {\Jl{Grav. Cosmol.}{#1}}
\def\GRG#1 {\Jl{Gen. Rel. Grav.}{#1}}
\def\JETF#1 {\Jl{Zh. Eksp. Teor. Fiz.}{#1}}
\def\JETP#1 {\Jl{Sov. Phys. JETP}{#1}}
\def\JHEP#1 {\Jl{JHEP}{#1}}
\def\JMP#1 {\Jl{J. Math. Phys.}{#1}}
\def\NPB#1 {\Jl{Nucl. Phys. B}{#1}}
\def\NP#1 {\Jl{Nucl. Phys.}{#1}}
\def\PLA#1 {\Jl{Phys. Lett. A}{#1}}
\def\PLB#1 {\Jl{Phys. Lett. B}{#1}}
\def\PRD#1 {\Jl{Phys. Rev. D}{#1}}
\def\PRL#1 {\Jl{Phys. Rev. Lett.}{#1}}

\def\lal{&&\nqq {}}

\def\beq{\begin{equation}}
\def\eeq{\end{equation}}
\def\bear{\begin{eqnarray}}
\def\bearr{\begin{eqnarray} \lal}
\def\ear{\end{eqnarray}}
\def\earn{\nonumber \end{eqnarray}}

\newcommand{\MI}{$\mathfrak{M}_1$}

\newcommand{\MIO}{$\mathfrak{M}^c_1$}

\newcommand{\Fig}[3]{%
\begin{center}
\parbox{8cm}{%
\refstepcounter{figure}\includegraphics[width=8cm,height=#2cm]{#1} \noindent Fig. \thefigure:\quad
#3}\end{center}}
\newcounter{strochka}
\newcommand{\stroka}[1]{\refstepcounter{strochka}\par\noindent\textsl{\Roman{spisok}.\arabic{strochka}}. \; \textsl{#1}}
\newcounter{spisok}
\newcommand{\spisok}[2]{%
\setcounter{strochka}{0}
\noindent \refstepcounter{spisok} \Roman{spisok}. \textbf{#1}
#2\hfill
}

\begin{document}
\thispagestyle{empty}
\twocolumn[

\vspace{1cm}

\Title{Single-field model of gravitational-scalar instability. II. Formation of black holes.\foom 1}

\Author{Yu. G. Ignat'ev}
    {Institute of Physics, Kazan Federal University, Kremlyovskaya str., 16A, Kazan, 420008, Russia}


\Abstract
 {On the basis of the previously formulated mathematical model of a statistical system with scalar interaction of fermions and the theory of gravitational-scalar instability of a cosmological model based on a one-component statistical system of scalarly charged degenerate fermions (\MIO models), the possibility of the formation of black holes in the early Universe using the mechanism of gravitational-scalar instability, which ensures the exponential growth of perturbations. The evolution of spherical masses in the \MIO model, as well as the evolution of black holes with allowance for their evaporation, is studied. The argumentation of the possibility of the formation of black holes in the early Universe with the help of the proposed mechanism is given, and a numerical model is constructed that confirms this argumentation. The range of parameters of the \MIO model, which ensures the growth of black hole masses in the early Universe up to $10^4\div10^6 M_\odot$, is identified.
}
\bigskip

] 
\section{Introduction}
In the first part \cite{GC_1part} of the Author's article, the development of gravitational-scalar instability in a one-component cosmological system of scalarly charged degenerate fermions with a classical scalar Higgs field, which we designated as the \MIO model, was studied in detail. The second part of the work is devoted to the study of the possibility of the formation of black holes in the early Universe using the mechanism of gravitational - scalar instability. At the same time, we will keep in mind the main goal of our study - to study the possibility of the formation of supermassive nuclei with masses\footnote{Recall that we use the Planck system of units $\hbar=G=c=1$. In this system $m_{pl}=1$, $M_\odot\approx10^{38}$. In the CGSE system $m_{pl}\approx 2\cdot 10^{-5}$g.}
\begin{equation}\label{M_nc}
M_{nc}\sim 10^4\div 10^6 M_\odot\approx 10^{42}\div10^{44}m_{pl}
\end{equation}
in the early Universe at redshifts $z\gtrsim 7$ corresponding to the age of the Universe $0.65\div 1\cdot10^9$ years (see, for example, \cite{SMBH1}, \cite{SMBH2}, \cite{Fan}) ). In what follows, we will focus on the lower estimate of this time, assuming
\begin{equation}\label{t_nc}
t_{nc}\sim 0.6\cdot10^9\mbox{years}\approx  3.8\cdot10^{59}t_{pl}.
\end{equation}

Let us single out some facts established in the previous part of the article \cite{GC_1part}, which are necessary in this part, taking into account the fact that we intend to further investigate cosmological models \MIO\ with an infinite future, which correspond to non-negative values of the cosmological constant
\begin{equation}\label{L>0}
\Lambda\geqslant0.
\end{equation}
Let us single out some facts established in the previous part of the article \cite{GC_1part}, which are necessary in this part, taking into account the fact that we intend to further investigate cosmological models \MIO\ with an infinite future, which correspond to non-negative values of the cosmological constant
\spisok{Basic provisions \cite{GC_1part}.}{
There are 3 types of instability\footnote{For $\Lambda<0$, there is one more type of instability, the narrow-band one, which we will not consider here.}, corresponding to different values of the scalar charge $e$ and the cosmological constant $\Lambda $.
\stroka{First type (2nd in \cite{GC_1part}); \emph{broadband instability} -- long phase with approximately constant $\gamma^\pm_-\approx \mathrm{Const}$ at $e<10^{-5},\ \Lambda<10^{-5} $; }
\stroka{The second type (3rd in \cite{GC_1part}) is \emph{quasi-periodic instability}: periodic alternation of pauses with a trapezoidal increment function $\gamma^-_-(t)$ and pauses with zero value $\gamma(t )=0$ (and the value of the increment maximum decreasing with time) for $e>5\cdot10^{-5}$, $\Lambda\gtrsim$ $10^{-5}$;}
\stroka{The third type (4th in \cite{GC_1part}) is \emph{aperiodic instability}: aperiodic phase alternation of instability with a constant maximum value $\gamma^-_-(t)$ at $e>10^{-5 }$, $\Lambda\gg 10^{-5}$.}
\stroka{\label{Bas2}
The value of the wave number $n$ affects the development of instability only at its early stages: as $n$ increases, the beginning of the unstable phase shifts to later times.}
}

Note that in the first part of the article, to simplify the study, we fixed the following parameters of the \MIO\ model: $\alpha=1$; $m=1$; $\pi_c=0.1$, and fixed the initial conditions $\Phi(0)=1,\ \dot{\Phi}(0)=0 $. In this part of the article, we will keep this fixation in order to avoid too cumbersome and obscure presentation of the research results. At the same time, we will assume that the main goal of this study is to identify the fundamental possibility of the formation of masses of the order \eqref{M_nc} as a result of the development of gravitational-scalar instability. We intend to work on finer tuning of the model, including the determination of the real parameters of the \MIO\ model, which ensure the generation of masses \eqref{M_nc}, in the future.

\section{Evolution of the reduced effective mass of the perturbed region}
\subsection{Background state for the cosmological model \MIO}
Let us write down the equations of the background spatially flat cosmological model necessary in what follows for a one-component system of scalarly charged degenerate fermions \cite{TMF_Ign_Ign21} (see details in \cite{GC_1part}, ($\xi=\ln a$, $\xi(0)=0$, $\pi_c\equiv\pi_{(z)}(0)$ is the Fermi momentum):
\begin{equation}\label{ap}
a(t)\pi_{(a)}(t)=\mathrm{Const}.
\end{equation}
\begin{eqnarray}
\label{dxi/dt-dPhi_Phi}
\dot{\xi}=H;\qquad \dot{\Phi}=Z;\\
\label{dH/dt_M1}
\dot{H}=-\frac{Z^2}{2}-\frac{4\mathrm{e}^{-3\xi}}{3\pi}
\pi_c^3\sqrt{\pi_c^2\mathrm{e}^{-2\xi}+e^2\Phi^2};
\end{eqnarray}
\begin{eqnarray}
\label{dZ/dt_M1}
\dot{Z}=-3HZ-m^2\Phi+\alpha\Phi^3-\nonumber\\[6pt]
\frac{4e^2\pi_c\mathrm{e}^{-\xi}}{\pi}\Phi\sqrt{\pi^2_c \mathrm{e}^{-2\xi}+e^2\Phi^2}+\nonumber\\
\frac{4e^4}{\pi}\Phi^3\ln\biggl(\frac{\pi_c\mathrm{e}^{-\xi}+\sqrt{\pi^2_c \mathrm{e}^{-2\xi}+e^2\Phi^2}}{|e\Phi|} \biggr);
\end{eqnarray}
The system of equations \eqref{dxi/dt-dPhi_Phi} -- \eqref{dZ/dt_M1} has as its first integral the total energy integral \cite{TMF_Ign_Ign21}
\begin{eqnarray}
3H^2-\Lambda+\frac{Z^2}{2}+\frac{z^2}{2}-\frac{m^2\Phi^2}{2}+\frac{\alpha\Phi^4}{4} \nonumber\\
-\frac{e^{-\xi}}{\pi}\pi_c\sqrt{\pi_c^2\mathrm{e}^{-2\xi}+e^2\Phi^2}\bigl(2\pi^2_c\mathrm{e}^{-2\xi}+e^2\Phi^2\bigr)\nonumber\\
\label{SurfEinst_M1}
+\frac{e^4\Phi^4}{\pi}\ln\biggl(\frac{\pi_c\mathrm{e}^{-\xi}+\sqrt{\pi^2_c \mathrm{e}^{-2\xi}+e^2\Phi^2}}{|e\Phi|} \biggr)=0.
\end{eqnarray}

Let us introduce the invariant characteristic of the unperturbed cosmological model, necessary in what follows, \emph{invariant cosmological acceleration}
\begin{equation}\label{Omega}
\Omega=1+\frac{\dot{H}}{H^2}.
\end{equation}
\subsection{Reduced mass and total charge of the perturbed region \MI}
Since the Einstein equation for the $^4_4$ components is the first integral of the complete dynamical system describing the unperturbed cosmological model, namely, the total energy integral, the effective energy density of the cosmological system, $\mathcal{E}_{eff}$, is determined from this integral as follows (see \cite{TMF_Ign_Ign21}):
\begin{equation}\label{E_eff}
\mathcal{E}_{eff}(t)=3H^2(t).
\end{equation}
Calculate \emph{reduced effective mass of the perturbation area} $M(n,t)$ -- the total energy - the mass enclosed in a sphere with a radius equal to the length of the perturbation $\lambda(t)=a(t)/n\equiv \mathrm {e}^{\xi(t)} /n$:
\begin{equation}\label{M_n}
M(n,t)=\frac{4\pi}{3}\lambda^3(t)\mathcal{E}_{eff}(t)=\frac{4\pi}{n^3}H^2(t)\mathrm{e}^{3\xi(t)}.
\end{equation}

Further, the total charge $Q(n,t)$ of the cor\-res\-ponding perturbation region is determined by integrating over the perturbation volume the scalar charge density $\rho(t)=e\pi_c^3/\pi^2a^3$
\begin{equation}\label{Q_n}
Q(n,t)= \int \rho dV=\frac{4}{3\pi}\frac{e \pi^3_c}{n^3}\equiv\frac{Q_0}{n^3}
\end{equation}
and does not depend on time.

Further, since the condition for WKB ap\-p\-li\-ca\-bi\-lity in the model under study
\begin{equation}\label{WKB}
n\eta=\int\limits_{t_0}^t \frac{dt}{a(t)}\gg1
\end{equation}
holds even for $n\gtrsim 10^{-2}$ (see \cite{GC_1part}), in what follows we assume $n=1$ in the formulas \eqref{M_n} and \eqref{Q_n}. Thus, we enter:
\begin{eqnarray}\label{M0}
M_0(t)=4\pi H^2(t)\mathrm{e}^{3\xi(t)};\quad Q_0=\frac{4}{3\pi}e \pi^3_c.
\end{eqnarray}

Note that at the stage of inflation ($H=H_0=\mathrm{Const}$, $\Omega=1$), according to \eqref{M_n}, the reduced perturbation mass grows exponentially
\begin{equation}\label{M_n-H0}
M(n,t)=\frac{4\pi}{n^3}H^2_0\mathrm{e}^{3H_0 t}.
\end{equation}

\subsection{Numerical modeling of the evolution of the perturbation mass}
Let us first find out how the model parameters $\Lambda$ and $e$ affect the evolution of the perturbation mass. On Fig. \ref{ignatev1} graphs of the evolution of the mass $M_0$ \eqref{M0} depending on the values of the cosmological constant at a fixed value of the scalar charge are presented.
\Fig{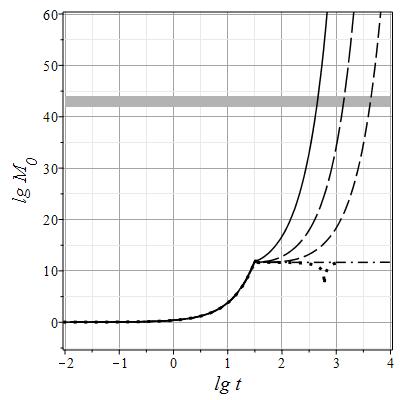}{7}{\label{ignatev1}Dependence of the mass of the perturbed region $M_0(t)$ \eqref{M0} on the value of the cosmological constant $\Lambda$ for $e=10^{-6}$: solid line -- $\Lambda=0.01$; long - dashed line -- $\Lambda=0.001$; dashed line -- $\Lambda=0.0001$, dash-dotted line -- $\Lambda=0$; dashed -- $\Lambda=-10^{-5}$.  The gray bar corresponds to the required mass values \eqref{M_nc}.}

As can be seen from these plots, firstly, in models with zero and negative values of the cosmological constant, the perturbation mass ceases to grow with time and tends to a constant value $M_0\sim 10^{11}\ m_{pl}$, which is much less than what we need. In this regard, in what follows we will consider models only with a positive value of the cosmological constant $\Lambda\geqslant0$. Secondly, as the cosmological constant $\Lambda$ increases, the required mass values are reached at earlier times of evolution.

Next, in Fig. \ref{ignatev2} graphs of the evolution of the mass $M_0$ \eqref{M0} depending on the value of the scalar charge at a fixed value of the cosmological constant are presented.

\Fig{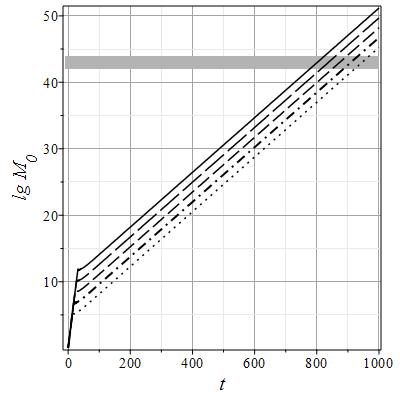}{7}{\label{ignatev2}Dependence of the mass of the perturbed region $M_0(t)$ \eqref{M0} on the value of the scalar charge $e$ at $\Lambda=0.003$: solid line -- $e=10^{-6}$; long - dashed line -- $e=10^{-5}$; dashed line -- $e=10^{-4}$, dashed line -- $e=10^{-3}$, dotted line -- $e=10^{-2}$. The gray bar corresponds to the required mass values \eqref{M_nc}.}
\Fig{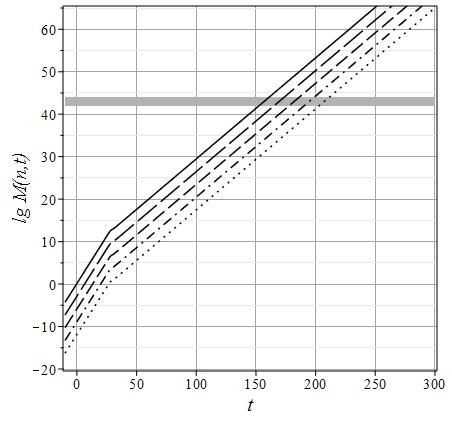}{7}{\label{ignatev3}Dependence of the mass of the perturbed region $M(n,t)$ \eqref{M_n} on the value of the wave number at $\Lambda=0.1$ and $e=10^{-5}$: dashed line -- $n=1$, dash - dotted -- $n=10$; solid line -- $n=100$. The gray bar corresponds to the required mass values \eqref{M_nc}. }
It follows from these graphs that with an increase in the scalar charge $e$, the mass values we need are reached at later times of evolution.

On Fig. \ref{ignatev3} plots of the evolution of the mass $M(n,t)$ \eqref{M_n} as a function of the wavenumber $n$ for fixed values of $\Lambda$ and $e$ are shown. As can be seen from these graphs, with an increase in $n$, the mass values we need are reached at later times of evolution. Thus, in order to achieve the required values \eqref{M_nc} of black hole masses during their evolution, small values of the scalar charges $e<10^{-5}$ and small wave numbers $n\lesssim 1$ are required at the same time sufficiently large values cosmological constant $\Lambda\gtrsim 10^{-5}$.

\section{Formation of black holes in\newline unstable perturbation modes}
\subsection{Why gravitational-scalar instability can lead to the formation of black holes?}
As can be seen from the previous one, we consider the instability of plane perturbations propagating in a given direction $\mathbf{n}$
\begin{equation}\label{petrurb}
\displaystyle f(\mathbf{r},t,\mathbf{n})=\tilde{f}(\mathbf{n},t)\mathrm{e}^{i\mathbf{nr}+\int u(\mathbf{n},t)dt/a(t)},
\end{equation}
where $\tilde{f}(\mathbf{n},t)$ are perturbation amplitudes slowly varying with time, and $u(\mathbf{n},t)$ are slowly varying with time, but large in absolute magnitude, eikonal functions.

How can these flat perturbations be transformed into spherically symmetric masses and then into black holes?

First, we note that, due to the macroscopic isotropy of the Universe, the per\-tur\-ba\-tions \eqref{petrurb} must also be distributed isotropically, i.e., must be averaged by some spherically symmetric in the space $\mathbb{N}_3=\{\mathbf{n}\}$ distribution function $F(\mathbf{n})$ over two-dimensional angles on the sphere, as well as over the length of the wave vector $|\mathbf{n}|$, i.e.,
over the spectrum of per\-tur\-ba\-tions\footnote{see see \cite{TMF_Ign_20}} for details. But why, then, in the standard theory of the formation of the large-scale structure of the Universe (see,
for example, \cite{Zeld}), do flat per\-tur\-ba\-tions, the so-called ``pancakes'', ultimately survive? First, for the formation of pancakes, a sufficiently long phase of the dusty state of matter
is necessary for the transformation of slowly growing ($\sim t^{2/3}$) per\-tur\-ba\-tions into nonlinear shock waves. Secondly, the per\-tur\-ba\-tion mode with the largest amplitude of one random direction
$\mathbf{n}$ survives due to the motion of dust particles that do not interact with each other. It is the dustiness factor of matter that leads to the formation of pancakes.

In the case of gravitational-scalar instability, the main component of matter is the scalar field, the dynamics of which is fundamentally different from the dynamics of non-interacting dust particles. Due to the following three factors, in our opinion, in the case of the development of gravitational-scalar instability, it is precisely spherically-symmetric mass distributions that develop under suitable conditions (more on this below) into black holes:\\

\spisok{Three factors}{
\stroka{Isotropic nature of perturbations.}
\stroka{Dependence of the beginning of the instability phase of the per\-tur\-ba\-tion mode on the wavenumber (basic provisions, I.\ref{Bas2}).\label{fac2}}
\stroka{An exponentially fast increase in per\-tur\-ba\-tions, due to which they quickly become non-linear.}
}

\vspace{6pt}
Due to the isotropy of the distribution of per\-tur\-ba\-tions, all per\-tur\-ba\-tions with the same wave number begin to grow simultaneously, regardless of the direction of the wave vector. In this case, per\-tur\-ba\-tions with small values of $n\sim1$ grow earlier. Further, due to the earlier instability of long-wave disturbances, their rapid growth and transition to the nonlinear stage, it is they who survive, forming spherically symmetric objects. The non-linear nature of per\-tur\-ba\-tions under suitable conditions should lead to the formation of black holes.

In addition, as will be seen below, there is also a \emph{fourth factor} that contributes to the survival of perturbation modes with a small wavenumber.

\subsection{The dynamics of the formation of black holes, taking into account their evaporation}
Let us now find out the fundamental possibility of the formation of black holes in the perturbation mode with the wavenumber $n$. As applied to the case of spherical collapse, the condition for the formation of a black hole is $R\leqslant 2r_g$, where $R$ is the radius of the collapsing object, $r_g=2M$ is its gravitational radius. Assuming the mass $M(n,t)$ \eqref{M_n} as the mass $M$ and the perturbation wavelength $\lambda(t)=a(t)/n$ used above as its radius, we obtain the necessary collapse condition perturbations with wavenumber $n$:
\begin{eqnarray}\label{Pi}
\frac{2M(n,t)}{\lambda(t)}-1>0\Rightarrow 
8\pi\frac{H^2}{n^2}\mathrm{e}^{2\xi(t)}-1>0.
\end{eqnarray}
The time of formation of a black hole, $t_g$, in the perturbation mode $n$ is then determined from the equation
\begin{equation}\label{t_n}
\displaystyle 8\pi\frac{H^2(t_g)}{n^2}\mathrm{e}^{2\xi(t_g)}=1.
\end{equation}
\Fig{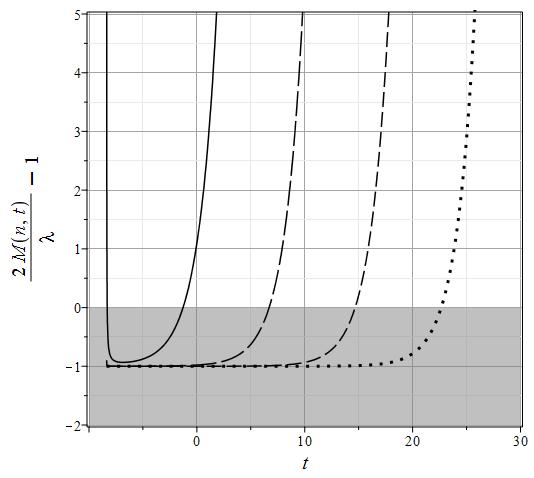}{7}{\label{ignatev4}Evolution of the black hole formation region (light part of the graph) depending on the wave number $n$ of perturbation for the parameters $\textbf{P}=[[1,1,10^{-6},0.1],10^{-6} ]$: solid line -- $n=1$; long - dashed line -- $n=10$; dashed line -- $n=100$, dotted line -- $n=1000$.}

On Fig. \ref{ignatev4} this time corresponds to the moment when the $2M(n,t)/\lambda(t)-1$ graph lines leave the gray area. Note that this value of time corresponds to the perturbation wavelength
\begin{equation}\label{L_g}
\lambda_g=\sqrt{\frac{3}{8\pi}}\frac{1}{H(t_g)}.
\end{equation}
In turn, this wavelength corresponds to a wave number $n_g$
\begin{equation}\label{n_g}
n_g=\sqrt{\frac{8\pi}{3}}H(t_g)\mathrm{e}^{\xi(t_g)},
\end{equation}
 such that all perturbations with $n\leqslant n_g$ collapse.

Commenting on the graphs in Fig. \ref{ignatev4}, note that, as it turns out, in addition to the above three factors that contribute to the formation of black holes in unstable perturbation modes, there is also a \emph{fourth factor}: -- black holes corresponding to large values of wave numbers can be formed in as a result of evolution much later than short-wavelength ones. Due to this factor, the effect of the factor is significantly enhanced.
II.\ref{fac2}, which contributes to the survival of the perturbation mode with a small wavenumber $n\sim1$\footnote{Under the condition that this mode is unstable.}. Note that this fourth factor is most directly related to the mechanism of black hole formation.

Taking into account the possibility of early formation of black holes in perturbation modes, we must also take into account the possible destruction of early black holes in the process of their evaporation. Let us compose the balance equation for the mass $M(t)$ of the black hole formed by the perturbation mode $n$, taking into account the process of black hole evaporation:
\begin{equation}\label{dM}
\frac{dM}{dt}=\frac{d M(n,t)}{d t}+\frac{d M_{e}}{dt},
\end{equation}
where
\begin{equation}\label{dM_em}
\frac{d M_{e}}{dt}=-\frac{1}{15\cdot2^{10}\pi}\frac{1}{M^2}\equiv -\frac{1}{\beta M^2}
\end{equation}
is the rate of evaporation of a black hole of mass $M$ (see, for example, \cite{Wald}). Differentiating with respect to time the expression for the perturbation mass \eqref{M_n}, we find, taking into account the definition of cosmological acceleration \eqref{Omega}
\begin{equation}\label{dM_n}
\frac{d M(n,t)}{dt}=H M(n,t)[3+2(\Omega-1)].
\end{equation}
Thus, we finally obtain the differential equation for the balance of the mass of a black hole formed by the perturbation mode $n$:
\begin{equation}\label{dM/dt}
\!\!\frac{dM}{dt}+\frac{1}{\beta M^2}=\frac{4\pi}{n^3}H^3(t)\mathrm{e}^{3\xi(t)}[3+2(\Omega(t)-1)].
\end{equation}
The equation \eqref{dM/dt} is an inhomogeneous first order differential equation. This equation must be solved with the initial condition
\begin{equation}\label{IC}
M(t_g)=M(t_g,n)
\end{equation}
and functions $\xi(t)$, $H(t)$, $\Omega(t)$ defined by the system of background equations \eqref{dxi/dt-dPhi_Phi} -- \eqref{dZ/dt_M1}.

On Fig. \ref{ignatev5} -- \ref{ignatev6} shows the results of numerical integration of the equation \eqref{dM/dt} together with the system \eqref{dxi/dt-dPhi_Phi} -- \eqref{dZ/dt_M1} and the initial condition \eqref {IC} for model parameters \MIO:
\begin{equation}\label{Pars1}
\mathbf{P_1}=[[1,1,10^{-7},0.1],3\cdot10^{-3}].
\end{equation}
\Fig{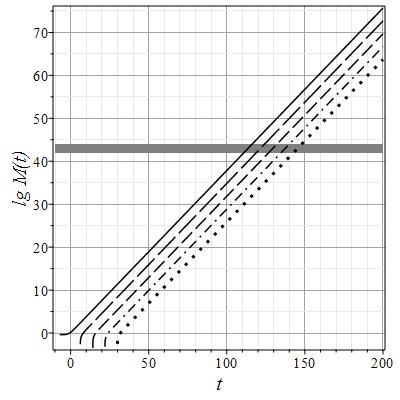}{7}{\label{ignatev5}The evolution of the black hole mass $M(t)$, taking into account the process of perturbed evaporation, depending on the value of the wave number at the parameters \eqref{Pars1}: solid line -- $n=1$, long-dashed line -- $n=10$; dashed line -- $n=100$; dash - dashed -- $n=1000$; dashed --$n=10000$. The gray bar corresponds to the required mass values \eqref{M_nc}. }

At the same time, in Fig. \ref{ignatev6} shows the early stages of the evolution of black hole masses. As expected, the deviation from the exponential de\-pen\-dence \eqref{M_n-H0}\footnote{linear for the logarithmic function}, which is typical for the plots in Fig. \ref{ignatev1} -- \ref{ignatev3}, manifests itself only at the earliest times $t\lesssim 20\ t_{pl}$, corresponding to the masses $M\lesssim 10 m_{pl}$, for which evaporation losses are significant.

\section{Numerical modeling of insta\-bi\-lity development}
So, we found out that the masses of fermion-scalar black holes of the order $M_{nc}$ \eqref{M_nc} can be achieved in our \MIO\ model only for positive values of the cosmological constant and small values of the scalar charge of fermions. In this case, the best results are given by models with a larger value of $\Lambda$. In this regard, let us consider in detail the case of parameters \eqref{Pars1}.

We are primarily interested in whether a suffi\-ci\-ently strong instability arises in a model with parameters \eqref{Pars1} for sufficiently small times and sufficiently small values of the wavenumber $n$ cor\-res\-pon\-ding to the plots in Fig. \ref{ignatev5} -- \ref{ignatev6}. In this model, the initial singularity corresponds to the time $t_0\approx-8.3280906$.
\Fig{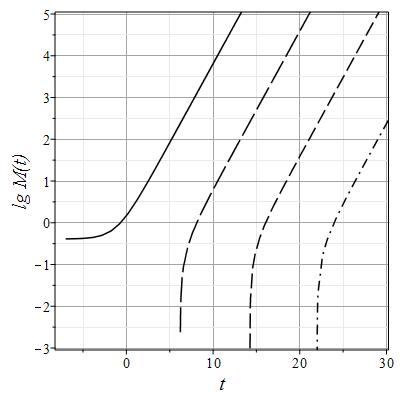}{7}{\label{ignatev6} Evolution of the black hole mass $M(t)$, taking into account the process of evaporation perturbed at early stages, depending on the value of the wave number at the parameters \eqref{Pars1}.}
\Fig{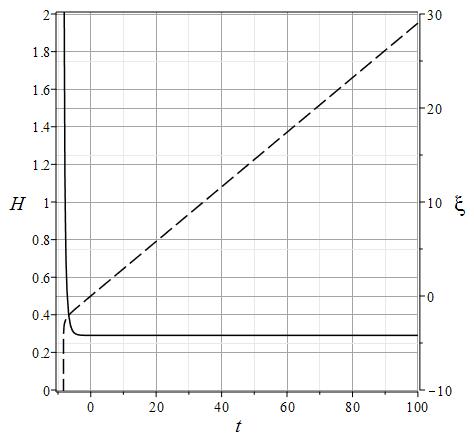}{7}{\label{ignatev7} Evolution of the scale function $\xi(t)$ (dashed line) and the Hubble parameter $H(t)$ (solid line) at the parameters \eqref{Pars1}.}
On Fig. \ref{ignatev7} shows the evolution of the scale function $\xi(t)$ and the Hubble parameter $H(t)$. By the time $t\backsimeq 0$, the model enters the inflationary expansion mode. At the same time, the scalar field potential remains constant $\Phi=1$, which corresponds to the stationary point of the Higgs vacuum scalar field dynamical system (see \cite{Ignat21_TMP}).

On Fig. \ref{ignatev8} plots of the evolution of the oscillation frequencies $\omega^\pm_-$ and growth increments of the per\-tur\-ba\-tion amplitude $\gamma^\pm_-$ for the perturbation mode with wavenumber $n=1$ are shown.
\Fig{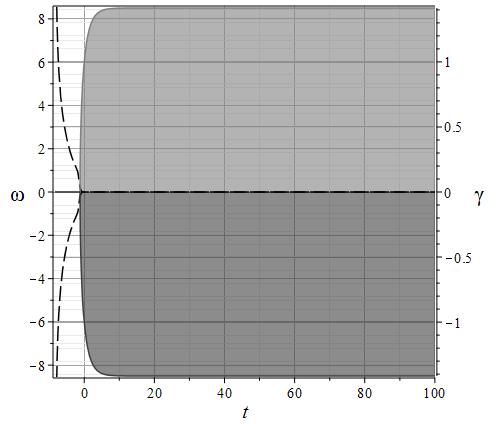}{7}{\label{ignatev8} The evolution of oscillation frequencies $\omega^\pm_-$ (dashed lines) and growth increments of the disturbance amplitude $\gamma^\pm_-$ (borders of light gray $\gamma^+_-$ and dark gray $\gamma^ -_-$ areas) with the parameters \eqref{Pars1} and $n=1$.}
Thus, for the parameters \eqref{Pars1} and $n=1$, the perturbation mode $(^+_-)$ becomes growing at $t\gtrsim0$, while the mode $(-_-)$ becomes damped, while an instability of the broadband type arises with $|\gamma^\pm_-|\approx 1.4$, which corresponds to standing growing and damping perturbations ($\omega^\pm_-=0$). At the same time, $\gamma^\pm_+=0$ (Fig. \ref{ignatev9}), which corresponds to traveling undamped waves with frequency decreasing with time (on the interval $t\in[0.50]$ of frequency $ \omega^\pm_+$ fall by 6 orders of magnitude.

Comparison of graphs in Fig. \ref{ignatev8} and Fig. \ref{ignatev5} -- \ref{ignatev6} shows that the beginning of the unstable phase for the $(^+_-)$ $n=1$ mode corresponds to the formation of a black hole with an initial mass $M(t_g)=1$, which during the time $\Delta t=10$ grows up to $10^4$ ($m_{pl}$) and then grows exponentially with time. Note that, in this case, the amplitude of this perturbation mode also grows according to the exponential law
\[\tilde{f}^+_-(t,1)\sim \mathrm{e}^{\gamma^+_- t}\approx \mathrm{e}^{1.4 t},\]
increasing $\exp(14)\approx 10^6$ times over the time interval $\Delta t=10$, which ensures the required increase in the black hole mass with a large margin.
\Fig{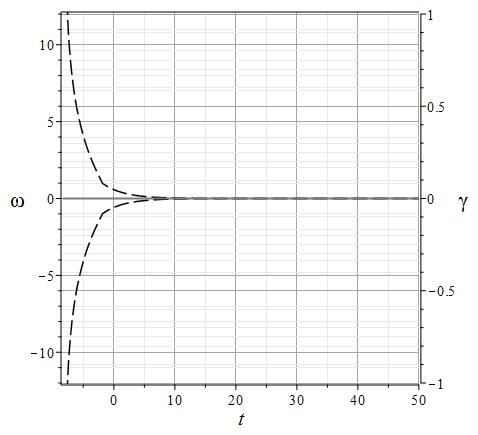}{7}{\label{ignatev9} The evolution of oscillation frequencies $\omega^\pm_+$ (dashed lines) and growth increments of the disturbance amplitude $\gamma^\pm_-$ are solid lines for the parameters \eqref{Pars1} and $n=1$.}
\section*{Conclusion}
Let us briefly list the main results of this part of the paper.\\

\spisok{Results:}{
\stroka{Based on the system of equations of the background cosmological model \MIO\, the evolution of the mass of matter $M(n,t)$, located in a sphere of radius of the perturbation wavelength with a given wavenumber $n$, is studied.
}
\stroka{The evolution of the black hole mass in the \MIO\ cosmological model is modeled taking into account the effect of black hole evaporation.
}
\stroka{The range of parameters of the \MIO model, which provides the necessary increase in the masses of black holes up to experimental estimates of the masses $M_{nc}=10^4\div10^6 M_\odot$, of the hypothesized nuclei of supermassive black holes in the centers of quasars at redshifts $z\gtrsim 7 $. In this case, it turns out that the required mass values can be reached much earlier, at times of the order of hundreds of Planck times.
}
\stroka{The argumentation of the possibility of the formation of early black holes in the event of a gravitational-scalar instability is given.
}
\stroka{A numerical model of the evolution of the instability of gravitational 0 scalar perturbations is constructed, which supports the version of the early formation of black holes.
}
}

Thus, we can state that the mechanism of formation of early black holes in unstable modes of gravitational-scalar perturbations of scalarly charged fermions is a productive idea that needs to be developed further.

In particular, as noted above, it is necessary to study the influence of the fundamental parameters of the Higgs potential on this mechanism, as well as the initial conditions, which were fixed in this study to simplify the modeling. It is also necessary to elucidate the possible contribution of phantom fields to the process of instability development. It is also important for us to study the nonlinear model of spherical collapse in a medium of scalarly charged fermions. In addition, it is necessary to find a mechanism that ensures further cessation of the growth of black hole masses. It is possible that a more complete model, taking into account phantom fields and the corresponding charges, will be able to solve this problem. This possibility was noted in \cite{GC_21_2}.

\subsection*{Funding}

 This paper has been supported by the Kazan Federal University Strategic Academic Leadership Program.


\begin{thebibliography}{15}
%
\bibitem{GC_1part}
Yu.G. Ignat'ev, Gravit. Cosmol., \textbf{28:3} (2020) (in print).
%
%
\bibitem{SMBH1}
S. Gillessen, F. Eisenhauer, S. Trippe, T. Alexander, R. Genzel, F. Martins, T. Ott, Astrophys.J., \textbf{692} 1075 (2009);
arXiv:0810.4674 [astro-ph].
%
\bibitem{SMBH2}
Sheperd Doeleman, Jonathan Weintroub, Alan E.E. Rogers  et al., Nature, \textbf{455} 78 (2008); arXiv:0809.2442 [astro-ph].
%
\bibitem{Fan}
X. Fan,  A Barth, E. Banados, G. D. Rosa,  R. Decarli, A.-C. Eilers et al., Bulletin of the AAS, \textbf{51(3)} (2019).
%
%
\bibitem{TMF_Ign_Ign21}
Yu.G. Ignat'ev and D.Yu. Ignat'ev, Theoret. and Math. Phys., \textbf{209:1}  1437 (2021); arXiv:2111.00492 [gr-qc].
%
\bibitem{TMF_Ign_20}
Yu.G. Ignat'ev, Theoret. and Math. Phys., \textbf{203:3}   927 (2020); arXiv:2004.14865[gr-qc].
%
\bibitem{Zeld}
Ja.B. Zeldovich, I.D. Novikov, \emph{Structure and Evolution of Universe}, Nauka, Moskow (1975) (in Russian).
[2] A.D. Dolgov, Ja.B. Zeldovich, Uspekhi fiz.
%
\bibitem{Wald}
%
Robert M. Wald, \emph{General Relativity}. The University of Chicago Press, Chicago (1984).
%
%
\bibitem{Ignat21_TMP}
Yu.G. Ignat'ev, I.A. Kokh, Theoret. and Math. Phys, \textbf{207:1} 514 (2021); arXiv:2104.01054 [gr-qc]. 
%
%
\bibitem{GC_21_2}
Yu. G. Ignat'ev, Gravit. Cosmol., \textbf{27:1}, 36 (2021); arXiv:2103.13867 [gr-qc].
%
%
\end{thebibliography}
\end{document}